# Constructing the Optimal Solutions to the Undiscounted Continuous-Time Infinite Horizon Optimization Problems

Dapeng CAI [1,*] and Takashi Gyoshin NITTA [2]

[1] *Institute for Advanced Research, Nagoya University, Furo-cho, Chikusa-ku, Nagoya, 464-8601, Japan; cai@iar.nagoya-u.ac.jp*  [2] *Department of Mathematics, Faculty of Education, Mie University, Kurimamachiya 1577, Tsu, 514-8507, Japan; nitta@edu.mie-u.ac.jp*

**Abstract**

We aim to construct the optimal solutions to the undiscounted continuous-time infinite horizon optimization problems, the objective functionals of which may be unbounded. We identify the condition under which the limit of the solutions to the finite horizon problems is optimal for the infinite horizon problems under the overtaking criterion.

*Keywords*: Infinite horizon optimization; Discounting

*JEL Classification Number*: C61; E21; O41

---

[*] *Corresponding Author*. Tel/Fax: +81-52-788-6141.

## 1. Introduction

Global climate change, radioactive waste disposal, loss of biodiversity, and many other challenges we face today require policies that could "treat the welfare of future generations on a par with our own" (Stern, 2007, p. 35). However, this is not easy with an infinite horizon, as the infinite series of utility sequences in general will diverge when future values are not discounted.[1]

Ramsey (1928) avoids the problem that the sum of the objective function may not converge by formulating the problem as minimizing the deviation from a given reference curve, the "bliss level". A more general approach that uses the concept of the overtaking criterion was later introduced by Von Weizsächer (1965) and Atsumi (1965) and further refined by Gale (1967) and Brock (1970).[2] Cai and Nitta (2007, 2008) explicitly construct the optimal paths for the discrete-time infinite horizon optimization problems that may have unbounded objective functionals. Their approach does not require the assumption of the existence of an optimal path. They conclude that under a fairly general condition, the conjecture that "the limit of the solutions for the finite problems is optimal for the infinite horizon problem" is correct for the discrete-time problems under the overtaking criterion.[3] In this paper, we aim to extend their results to continuous-time infinite horizon optimization problems that may have unbounded objective functionals. We also demonstrate the applicability of the results by considering an example.

---

[1] Discounting avoids this problem, however, economists have long been scathing about the ethical and logical difficulties it generates (Ramsey, 1928; Pigou, 1932; Harrod, 1948; Solow, 1974; Heal, 1998; Weitzman, 1998; Stern, 2007).

[2] Under the assumption of the existence of a given reference curve, the existence and dynamical properties of the resultant optimal path have been considered in, for example, Michel (1990), Durán (2000), Kamihigashi (2001), and Le Van and Morhaim (2006).

## 2. The Model

We consider an economy that is composed of many identical households, each forming an immortal extended family. Given the planning horizon $T \in [0, \infty)$, the criterion for a social planner to judge the welfare of the representative household takes the form

$$\max_{c(t)} \int_0^T U(c(t))dt, \qquad (1)$$

where the instantaneous utility function $U: \mathbb{R}_+ \to \mathbb{R}_+$ is continuous, strictly increasing, strictly concave, and continuous differentiable. At each time, a representative household invests $k(t)$ to produce $f(k(t))$ amount of output. The production process is postulated as follows: $f: \mathbb{R}_+ \to \mathbb{R}_+$, $f$ is continuous, strictly increasing, weakly concave, and continuous differentiable, with $f(0) = 0$. The household chooses a path $c(t)$ that maximizes (1), which is subject to the budget constraint:

$$\dot{k}_t = f(c_t, k_t), \qquad (2)$$

given the initial capital stock, $k(0) \equiv k_0 > 0$. We consider a free terminal state problem.

---

[3] For the case when the discount factor is less than 1, Stokey and Lucas (1989) examine the conjecture by using recursive methods. Proving the conjecture involves establishing the legitimacy of interchanging the operators "max" and "$\lim_{T \to \infty}$".

A unique optimal solution to problem (1), subject to (2) and given the initial capital stock, $\{c_T(t), k_T(t)\}_{t=0}^{T}$, can be found readily by setting up the Hamiltonian. We proceed to extend the planning horizon to infinity. An immediate problem is that the infinite series $\int_0^\infty U(c(t))dt$ in general will diverge and the maximization of which may be meaningless in $\mathbb{R}$. In what follows, we identify the condition under which the limit of the solutions to the finite horizon problems is optimal among all attainable paths for the infinite horizon problems, under the overtaking criterion defined below.

**Definition 1.** $(c_1, k_1)$ and $(c_2, k_2)$ are two attainable paths. For all $U > 0$, $(c_2, k_2)$ overtakes $(c_1, k_1)$ if $\lim_{T \to \infty} \left( \int_0^T U(c_1(t))dt - \int_0^T U(c_2(t))dt \right) < 0$.

Following Brock's (1970) notion of *weak maximality*, the optimality criterion in this paper is defined as follows:

**Definition 2.** An attainable path $(c(t), k(t))$ is optimal if no other attainable path $(c_1, k_1)$ overtakes it: $\lim_{T \to \infty} \left( \int_0^T U(c_1(t))dt - \int_0^T U(c(t))dt \right) \leq 0$.

Denote $(c°(t), k°(t))$ as the limit of $\{c_T(t), k_T(t)\}_{t=0}^{T}$ when the planning horizon $T$ grows to infinity: $c°(t) \equiv \lim_{T\to\infty} c_T(t)$ and $k°(t) \equiv \lim_{T\to\infty} k_T(t)$.

**Theorem.** If $\lim_{T\to\infty} \left( \dfrac{\int_0^T (U(c°(t)) - U(c_T(t))) dt}{\int_0^T U(c°(t)) dt} \right) = 0$, then no other attainable path $(c(t), k(t))$ overtakes $(c°(t), k°(t))$.

**Proof.** For any attainable path $(c(t), k(t))$, we see that

$$\frac{\int_0^T U(c(t)) dt}{\int_0^T U(c°(t)) dt} = \frac{\int_0^T U(c(t)) dt}{\int_0^T U(c_T(t)) dt} \cdot \frac{\int_0^T U(c_T(t)) dt}{\int_0^T U(c°(t)) dt}.$$

Since $c_T(t)$ is the optimal path in $[0, T]$, we have $\dfrac{\int_0^T U(c(t)) dt}{\int_0^T U(c_T(t)) dt} \leq 1$. Moreover,

under the assumption $\lim_{T\to\infty} \left( \dfrac{\int_0^T (U(c°(t)) - U(c_T(t))) dt}{\int_0^T U(c°(t)) dt} \right) = 0$, we see that

$$\lim_{T\to\infty} \frac{\int_0^T U(c_T(t)) dt}{\int_0^T U(c°(t)) dt} = \lim_{T\to\infty} \left( \frac{\int_0^T (U(c_T(t)) - U(c°(t))) dt + \int_0^T (U(c°(t))) dt}{\int_0^T U(c°(t)) dt} \right)$$

$$= \lim_{T\to\infty} \left( \frac{\int_0^T (U(c_T(t)) - U(c°(t))) dt}{\int_0^T U(c°(t)) dt} \right) + 1 = 1.$$

From Lemma 1 in Cai and Nitta (2007)[4], we see that

$$\lim_{T\to\infty}\left(\frac{\int_0^T U(c(t))dt}{\int_0^T U(c_T(t))dt} \cdot \frac{\int_0^T U(c_T(t))dt}{\int_0^T U(c(t))dt}\right) = \lim_{T\to\infty}\left(\frac{\int_0^T U(c(t))dt}{\int_0^T U(c_T(t))dt}\right) \cdot \lim_{T\to\infty}\left(\frac{\int_0^T U(c_T(t))dt}{\int_0^T U(c(t))dt}\right)$$

$$= \lim_{T\to\infty}\left(\frac{\int_0^T U(c(t))dt}{\int_0^T U(c_T(t))dt}\right) \leq 1.$$

Again from Lemma 1 in Cai and Nitta (2007), we have

$$\lim_{T\to\infty}\left(\int_0^T U(c(t))dt - \int_0^T U(c^\circ(t))dt\right) = \lim_{T\to\infty}\left(1 - \frac{\int_0^T U(c(t))dt}{\int_0^T U(c^\circ(t))dt}\right)\lim_{T\to\infty}\int_0^T U(c^\circ(t))dt.$$

Because $\lim_{T\to\infty}\int_0^T U(c^\circ(t))dt > 0$ and $\lim_{T\to\infty}\left(\frac{\int_0^T U(c(t))dt}{\int_0^T U(c^\circ(t))dt}\right) \leq 1$, we see that

$$\lim_{T\to\infty}\left(\int_0^T U(c(t))dt - \int_0^T U(c^\circ(t))dt\right) \leq 0. \qquad \text{Q.E.D.}$$

## 3. An Example

Find the curve with the shortest distance from a given point to a given straight line.

---

[4] The lemma is as follows: Let $a_T$ and $b_T$, $T \in [0,\infty)$, be two sequences. If $\lim_{T\to\infty} a_T > 0$, $\lim_{T\to\infty} b_T > 0$, then $\lim_{T\to\infty}(ab)_T = \lim_{T\to\infty} a_T \cdot \lim_{T\to\infty} b_T$.

$$\text{Maximize } V = \int_0^\infty -\left(1+u^2\right)^{1/2} dt \qquad (3)$$

subject to $\dot{y} = u$, and $y(0) = A$, $y(T)$ free ($A, T$ given)

By setting up the Hamiltonian function of the finite time version of the problem, it is easy to show that for all $t \in [0, T]$, the optimal solutions are

$$\lambda_T^*(t) = 0, \qquad (4)$$

$$u_T^*(t) = 0, \qquad (5)$$

$$y_T^*(t) = A, \text{ since } \dot{y} = 0 \text{ and } y(0) = A. \qquad (6)$$

Let $u^\circ(t) \equiv \lim_{T \to \infty} u_T^*(t) = 0$, $y^\circ(t) \equiv \lim_{T \to \infty} y_T^*(t) = A$. It is easy to verify that the condition in Theorem is satisfied, and $(u^\circ(t), y^\circ(t))$ is the optimal path.

## 4. Concluding Remarks

In this paper, we extend Cai and Nitta (2007, 2008) to continuous-time problems. We show that the approach is readily applicable for models that satisfy a fairly general condition, with the path obtained by taking the limit of the solutions to the finite horizon problems being the optimum for the infinite horizon problem. For most examples, however, it may not be possible to explicitly check the condition in Theorem and study the resultant paths. In such cases, a numerical approach can be used to check the condition and to compute explicit solutions.

# References

Atsumi, H. 1965, Neoclassical growth and the efficient program of capital accumulation, Review of Economic Studies 32, 127–136.

Brock, W.A. 1970, On the existence of weakly maximal programmes in multi-sector economy, Review of Economic Studies 37, 275-280.

Cai, D. and G.T. Nitta, 2007, Treating the future equally: Solving undiscounted infinite horizon optimization problems, arXiv:math/0701371v4[math. OC].

Cai, D. and G.T. Nitta, 2008, Limit of the solutions for the finite horizon problems as the optimal solution to the infinite horizon optimization, mimeo. (available upon request)

Durán, J. 2000, On dynamic programming with unbounded returns, Economic Theory 15, 339-352.

Gale, D. 1967, On optimal development in a multi-sector economy, Review of Economic Studies 34, 1–18.

Harrod, R.F. 1948, Towards a Dynamic Economics. (Macmillan, London).

Heal, G. 1998, Valuing the Future: Economic Theory and Sustainability. (Columbia University Press, New York).

Kamihigashi, T. 2001, Necessity of transversality conditions for infinite horizon problems, Econometrica 69, 995-1012.

Le Van, C. and L. Morhaim, 2006, On optimal growth models when the discount factor is near 1 or equal to 1, International Journal of Economic Theory 2, 55-76.

Michel, P. 1990, Some clarifications on the transversality condition, Econometrica 58, 705-723.

Pigou, A.C. 1932, The Economics of Welfare. 4th ed. (Macmillan, London).

Ramsey, F.P. 1928, A mathematical theory of saving, Economic Journal 38, 543-559.

Stern, N. 2007, The Economics of Climate Change: The Stern Review. (Cambridge University Press, Cambridge).

Solow, R.M. 1974, Richard T. Ely lecture: The economics of resources or the resources of economics, American Economic Review 64, 1-14.

Stokey, N. and R. Lucas, 1989, Recursive Methods in Economic Dynamics. (Harvard University Press, Cambridge).

Von Weizäcker, C.C. 1965, Existence of optimal programs of accumulation for an infinite time horizon, Review of Economic Studies 32, 85–104.

Weitzman, M.L. 1998, Why the far-distant future should be discounted at its lowest possible rate, Journal of Environmental Economics and Management 36, 201-208.